\documentclass[12pt]{article}

\renewcommand{\bar}[1]{\overline{#1}}
\newcommand{\half}{{\frac{1}{2}}}
\newcommand{\threehalfs}{{\frac{3}{2}}}
\newcommand{\VEV}[1]{\left\langle{#1}\right\rangle}

\begin{document}

\begin{flushright}

SLAC-PUB-10200\\
October 2003\\
\end{flushright}

\bigskip

{\centerline{\Large \bf  Light-Front Hadron Dynamics}
\vspace{10pt}
\centerline{\Large \bf {and AdS/CFT Correspondence}
\footnote{\baselineskip=13pt
Work partially supported by the Department of Energy, contract
DE-AC03-76SF00515.}}

\vspace{15pt}

\centerline{\bf Stanley J. Brodsky
\footnote{\baselineskip=13pt
{E-mail: sjbth@slac.stanford.edu}}}
{\centerline{Stanford Linear Accelerator Center,}}
{\centerline{Stanford University, Stanford, California 94309, USA}}
\vspace{8pt}
\centerline{\bf Guy F. de T\'eramond
\footnote{\baselineskip=13pt
{E-mail: gdt@asterix.crnet.cr}}}
{\centerline{Universidad de Costa Rica, San Jos\'e, Costa Rica}}

\vspace{30pt}

\begin{abstract}

A remarkable consequence of the AdS/CFT correspondence is the
nonperturbative derivation of dimensional counting rules for hard
scattering processes.  Using string/gauge duality we derive the
QCD power behavior of light-front Fock-state hadronic
wavefunctions for hard scattering in the large-$r$ region of the
AdS space from the conformal isometries which determine the
scaling of string states as we approach the boundary from the
interior of AdS space. The nonperturbative scaling results are
obtained for spin-zero and spin-$\half$ hadrons and are extended
to include the orbital angular momentum dependence of the
constituents in the light-front Fock-expansion. The correspondence
with string states is considered for hadronic states of arbitrary
orbital angular momentum for a given hadron of spin $\leq 2$.
We examine the implications of the color configuration
of hadronic Fock-states for the QCD structure of scattering
amplitudes at large $N_C$. Quark interchange amplitudes emerge as the
dominant large $N_C$ scattering mechanisms for conformal QCD.

\end{abstract}

\centerline{
}



\vfill

\newpage

~~~Since the early identification of large $N_C$ conformal QCD
with the topological expansion of a string~\cite{'tHooft:1973jz},
the connection between string theories and the large-$N_C$ limit
of field theories has drawn major attention. However, string
theory is not consistent in a four dimensional flat space, and at
least one extra dimension has to be
introduced~\cite{Polyakov:1997tj}. It is only recently that a
precise correspondence has been established between quantum field
theories and string/M-theory on Anti-de Sitter spaces
(AdS)~\cite{Maldacena:1997re}, where strings live on the curved
geometry of the AdS space and the observables of the conformal
field theory are defined in the boundary of  AdS. The partition
function of the AdS theory is identified with the generating
functional of the boundary conformal gauge field theory, and
correlation functions are computed in the boundary of AdS where
the quantum field theory operators are
inserted~\cite{Gubser:1998bc, Witten:1998qj}. The AdS/CFT
conjecture is indeed a concrete realization of the holographic
principle\\~\cite{'tHooft:gx, Susskind:1994vu}. From the point of
view of string theory/field theory duality, excitations near the
boundary of the conformal AdS space correspond to states in the
field theory. Hard-scattering interactions occur in the large-$r$
region of the AdS space, so effectively we require the 't Hooft
parameter, the product of the string coupling and the number of
colors $g_s N_C$,  to be sufficiently large so that the AdS radius
$R = ({4 \pi g_s N_C})^{1/4} \alpha'^{1/2}$, with  $\alpha'$ the
string scale, is also large. We can map the string states degrees
of freedom to the QCD degrees of freedom at the boundary of the
AdS space~\cite{Aharony:1999ti}.

A remarkable consequence of the AdS/CFT correspondence is the
derivation~\cite{Polchinski:2001tt} of dimensional counting rules
for the leading power-law fall-off of hard exclusive
processes~\cite{Brodsky:1973kr, Matveev:ra}. The derivation from
supergravity/string theory in a  $AdS_5 \times X$ background does
not rely on perturbation theory, and thus is more general than
perturbative QCD analysis~\cite{Lepage:1980fj}.
Classical QCD with massless quarks is conformal.
Corrections from nonconformal effects in QCD due to
the inverse logarithmic running of the coupling are
moderate in the ultraviolet region.
Theoretical~\cite{vonSmekal:1997is,Zwanziger:2003cf,Howe:2002rb} and
phenomenological~\cite{Mattingly:ej,Brodsky:2002nb}  evidence is now
accumulating that the QCD coupling becomes constant at small virtuality; i.e.,
$\alpha_s(Q^2)$ develops an infrared fixed point.
Indeed,  QCD appears to be a nearly-conformal theory in the momentum
regimes accessible to experiment. The amplitudes of confining gauge
theories with superstring duals, in addition to the power-law hard
dependence at large momentum transfer, have Regge behavior at small
angles for different kinematic domains~\cite{Polchinski:2001tt}. The
String/Gauge duality for QCD in four dimensions can also be obtained from
M-theory in a specific Black Hole deformation of
$AdS^7 \times S^4$~\cite{Brower:2002er} with Regge behavior in the
near-forward limit~\cite{Boschi-Filho:2002zs}. The AdS/CFT correspondence
for gauge theories with dual
strings  has been used to describe deep inelastic scattering
~\cite{Polchinski:2002jw}, and the derivation of scaling laws in
hadronic processes has been considered
for exclusive processes, form factors and
structure functions~\cite{Andreev:2002aw}.

In this paper counting rules for light-front wavefunctions, the probability
amplitudes which relate the constituents degrees of freedom with the asymptotic
hadronic states, are obtained from the AdS/CFT correspondence for large momentum
transfer processes as we approach the boundary of the AdS space. The
nonperturbative scaling results are obtained for spin-zero and spin-${\half}$
hadrons. We are indeed limited to string states dual to hadrons of spin $0,
\half, 1$, $\threehalfs$ and $2$, since in string theory there are only fields of
spin $\leq 2$, and consequently all fields with higher spin have masses of order
of the string scale $(\alpha')^{-1/2}$. The counting rules for light-front
wavefunctions are also extended to include the orbital angular momentum
dependence of the constituents by computing the corresponding scaling dimension
from the conformal algebra of the generators and their action on the field
operators. We also discuss the AdS/CFT string/gauge theory duality for
Fock-states with orbital angular momentum by examining the Kaluza-Klein (KK)
modes in the compact manifolds required to match the 10-dimensional
critical-strings. Our goal is to use the AdS/CFT correspondence to constrain the
form of the light-front wavefunctions of hadrons in QCD in the phenomenologically
interesting near-conformal regime. Consequently we will evaluate the dual string
theory at finite 't Hooft coupling $g_s N_C$ and well defined values of  $\lambda
R$, where $\lambda$ is the characteristic energy of a KK excitation.

It is well known that the quark interchange amplitudes for fixed-angle exclusive
processes constitute a dominant  mechanism in the large 't Hooft
limit~\cite{'tHooft:1973jz} $N_C \to \infty$, keeping the product $g^2_{QCD}(\mu)
N_C$, with $g^2_{QCD} \sim g_s$,  large but fixed at some given scale $\mu$. The
quark interchange amplitudes are planar Feynman diagrams and are leading
amplitudes in a ${1/N_C}$ expansion, with other non-planar contributions being
suppressed by powers of $1/N_C$. This result is related to hadronic
duality~\cite{Harari:nn}, since in the large $N_C$ limit quark-interchange
amplitudes are topologically equivalent to duality diagrams which keep track of
the flow of quark quantum numbers; they represent string world-sheets with the
quark lines defining the boundaries. There is also a set of graphs corresponding
to multiple gluon exchange, where the gluons are connected to the same quark at
each hadron and are clearly planar. However, planarity is not sufficient to
ensure that a diagram is leading in the $1/N_C$ expansion. Multiple gluon
exchange planar diagrams are also considered, and it is shown that they are
suppressed relative to quark interchange.

The quark interchange amplitude for meson-meson
scattering scales as $1/N_C$ while other contributions scale faster as
$N_C \to \infty$. The case for baryons is more complicated
since the baryons have $N_C$ quarks in their valence state, and the
usual description  of baryon elastic scattering
~\cite{Witten:1979kh} gives a result  which diverges linearly with $N_C$
for quark exchange amplitudes. It will be shown below that if proper account is taken
of the color configuration in the normalization of the Fock components,
the antisymmetrization properties of the $SU(N_C)$ baryon wave
function leads to the same $N_C$ counting rules for baryon and  meson states.
This is quite satisfactory. For example, the sum of the photon attachments in a form
factor should give the total charge independent of $N_C$, regardless that we are
describing a meson or a baryon.
The string/field theory duality predictions for hard
scattering correspond precisely to the $N_C \to \infty$ limit,
which is also the large $g_S N_C$ limit. For large transverse
momentum the relevant scale is $\mu \sim Q,$ and quark exchange
interactions occur in the large-$r$ conformal region of the $AdS$
space.

In QCD the Hilbert space is constructed in terms of a
complete Fock basis of non interacting constituents at fixed light-front time,
where the amplitudes are $n$-parton light-front wave
functions $\psi_{n/h}$ corresponding to the expansion of color
singlet hadron states
\begin{equation}
\vert \Psi_h \rangle = \sum_n \psi_{n/h} \vert n \rangle .
\label{eq:Fockexp}
\end{equation}
The light-front Fock-state wavefunction provides a
frame-independent representation of relativistic composite systems
in QCD at the amplitude level, in terms of quark and gluon degrees
of freedom which carry the symmetries within the hadrons.

According to the AdS/CFT correspondence, a quantum gravity theory in AdS space defines a
conformal field theory on its boundary~\cite{Maldacena:1997re}, where a string state is
related with its corresponding renormalized four-dimensional quantum field
operator~\cite{Gubser:1998bc, Witten:1998qj}. We thus expect that for large values of the
't Hooft parameter $g_s N_C$, near-conformal QCD has an effective
dual string description. We will
show that the scaling properties of the $n$-parton light-front wave functions are obtained
from the string/gauge theory duality, and their power-law dependence follows from the
conformal dimension of the corresponding string state dual to hadron states. The zero-mode
states of strings can be treated classically, whereas the higher Fock components of Eq.
(\ref{eq:Fockexp}) are a manifestation of the quantum fluctuations of QCD.
Indeed, Hilbert space is defined in the boundary theory and not within the interior
of AdS space. Since
string theory predicts the scaling behavior of string states corresponding to different
numbers of constituents, we can apply this information to the higher Fock components of
physical hadrons and their corresponding interpolating operators. The correspondence is
particularly useful for exclusive reactions at large momentum which are controlled by Fock
components with a minimum number of constituents. We will also examine the correspondence
of hadron Fock states with non-zero orbital angular momentum and  string states with
Kaluza-Klein excitations in the compact space $X$. In this case the discussion is more
complex, since the dimensions of the states are in general not protected from acquiring
large values at strong coupling, and considerations about the chirality properties of
states are important.

To conserve Poincar\'e invariance
in four dimensional Minkowski space the metric should have the form
\begin{equation}
ds^2 = w(z)^2 [dx^2 - dz^2],
\label{eq:AdSz}
\end{equation}
where $dx^2 = \eta_{\mu \nu} dx^\mu dx^\nu$. Reparametrization
invariance allows us to factor out the warp factor $w(z)^2$. The
coordinates $x^\mu$ are Minkowski spacetime variables and $z$ labels
the new spatial dimension.  If the four-dimensional gauge theory
is conformal, it is invariant under the transformation $x \to
\rho x$. If $z \to \rho z$ is also an isometry of
(\ref{eq:AdSz}), then $w(z) = R/z$ and the space is a
five-dimensional Anti-de Sitter space, $AdS_5$, and the conformal
group $SO(2,4)$ in Minkowski space is also the group of symmetries
of $AdS_5$. The value $z = 0$ corresponds to the
four-dimensional boundary where the gauge theory is defined.
Introducing a radial coordinate $r = R^2/z$ we can express the
$AdS_5$ metric (\ref{eq:AdSz}) as
\begin{equation}
ds^2 = \frac{r^2}{R^2} \left( dt^2 -  d\vec x^2 \right) -
\frac{R^2}{r^2} dr^2  . \label{eq:AdSr}
\end{equation}

If the conformal gauge theory is dual to a critical string, a superstring
theory in 10 dimensions, spacetime is the product space $AdS_5$
with a five-dimensional manifold $X$.
In the original Maldacena conjecture~\cite{Maldacena:1997re} a correspondence was
established between Type IIB string - a closed string
with left-right asymmetric vibrational patterns -
compactified on $AdS_5 \times S^5$,
and ${\cal N} = 4$ large $N_C$ superconformal Yang-Mills theory at
the boundary, with the solution $R = ({4 \pi g_s N_C})^{1/4}
\alpha'^{1/2}$  for the radius of AdS and the
radius of the five-sphere. The theory with maximal
supersymmetry corresponds to the sphere.

The observables of the gauge theory are defined at the boundary of
the AdS space. The AdS metric implies that a distance $L_{\rm local}$,
measured in the local inertial coordinates in the full
space shrinks by a warp factor as one approaches the AdS boundary.
Thus the length at large-$r$ as observed in Minkowski space is
\begin{equation}
L_r = \left(\frac{R}{r}\right) ~ L_{\rm local}.
\end{equation}
Consequently, a local scattering state in the bulk with local
momentum $Q_{\rm local}$ as determined in the inertial frame is
shifted to the ultraviolet as measured by an observer in Minkowski
space
\begin{equation}
Q_r = \left(\frac{r}{R}\right) ~Q_{\rm local}.
\end{equation}
Since the characteristic energy scale in the ten-dimensional space
is $Q_{\rm local} \sim R^{-1}$, it follows that $ r \sim Q R^2$.
Thus large $Q$ pointlike hard scattering processes occur in the
large-r conformal region of the AdS space.

The geometry at small values of $r$ and the form of the compact
space $X$ depend on the nature of the gauge theory and the
mechanisms of conformal symmetry breaking. QCD is a nearly
conformal theory in the ultraviolet region and a confining theory
in the infrared with a mass gap  corresponding to the scale
$\Lambda_o$ of the lightest glueball scale. We thus expect that a
non-conformal metric will be manifest at $r \sim r_o = \Lambda_o ~
R^2$. However, we are interested in probing the wavefunction at
large transverse momentum of order $Q \sim r/R^2 \gg \Lambda_o.$
The interaction occurs in the $ r \sim Q R^2 \gg r_o$ region where
the conformal geometry (\ref{eq:AdSr}) is valid and where the
specific form of the compact space $X$ is not
relevant~\cite{Polchinski:2001tt}.

Coordinates on $AdS_5$ are  $x^\mu$ and $r$, and coordinates in the compact space
are labelled $y^a$. According to the AdS/CFT correspondence we expect hadronic
states at the boundary to have dual scattering states in the bulk. Consider first
a spin-zero hadron state with four momentum $P^\mu$ in four-dimensional Minkowski
space. The physical hadron in the boundary theory is expected to be  dual to a
string state represented by a 10-dimensional wavefunction $\Phi$, which is
identified with a spinless dilaton field~\cite{Polchinski:2001tt} satisfying the
Laplace equation in ten-dimensional curved spacetime. For large values of $ r \gg
r_o$ spacetime geometry is the product space $AdS \times X$. The field $\Phi$ is
expanded in terms of a complete set of normalized eigenfunctions $\phi_l$  of the
five-dimensional Laplace operator, with proper boundary conditions on the compact
space $X$ and eigenvalues $\lambda_l^2$. We write $\Phi(x, r, y) = \sum_l
\Psi_l(x, r) \phi_l(y)$.  From the point of view of ten-dimensional physics
$\Phi$ is a string zero-mode of the ten-dimensional Laplace equation. However, we
can still have a complete set of excited states corresponding to the Kaluza-Klein
modes yet having zero modes in ten dimensions. Nonzero eigenvalues on $X$ are of
the order of the inverse of the radius of the compact space $X$. For each
eigenvalue $\lambda_l$ in the compact manifold $X$, the field $\Psi$ satisfies a
d+1 dimensional scalar equation in the warped geometry of $AdS_{d+1}$ space with
the linearly independent solutions
\begin{equation}
\Psi(x,r) = C e^{-i P \cdot x} r^{-\frac{d}{2}}
\left\{
\begin{array}{l}
 J_\alpha\left(\frac{{\cal M} R^2}{r}\right)\\[.5ex]
 N_\alpha\left(\frac{{\cal M} R^2}{r}\right)
\end{array}
\right. \ ,
\label{eq:scalarAdS}
\end{equation}
where  $\alpha^2  = (d/2)^2 + (\lambda R)^2$ and
${\cal M}$ is the hadronic invariant mass, $P_\mu P^\mu = {\cal M}^2$.
The dominant contribution at
large-$r$ is $\Psi(r) \sim r^{-\Delta}$,
where
\begin{equation}
\Delta = \half \left( d  + \sqrt{  d^2 +
 4 \lambda^2 R^2 }~\right) \label{eq:DeltaM}
\end{equation}
is the conformal dimension for a scalar field in $AdS_{d+1}$. At large relative transverse
momentum, $Q \sim r/ R^2$,  $\Psi$ has the scaling
behavior $\Psi(Q) \sim Q^{- \Delta}$
in the large-$r$ region of AdS space.
The limiting values of the operators in the bulk are the
Heisenberg field operators of the gauge field
theory~\cite{Maldacena:1997re} specified by the boundary
conditions~\cite{Gubser:1998bc, Witten:1998qj}
\begin{equation}
\lim_{r \to \infty} \Psi(x,r) =  r^{-\Delta} \Psi_o(x),
\label{eq:bc}
\end{equation}
where $\Psi_o(x)$ is the corresponding renormalized operator of
the quantum field theory~\cite{ref:GF}.
We shall first establish the correspondence
with the QCD partonic degrees of freedom in the Fock expansion of hadronic
quantum states in the AdS boundary for the case $\lambda = 0$, corresponding to a
zero mode in $X$ space.
We will return to the excited modes on $X$, when discussing the extension of our
results to include the orbital angular momentum of the Fock components.

The basic constituents in QCD appear from the light-front quantization of the
excitations of the dynamical fields, expanded in terms of creation
and annihilation operators on the transverse plane with
coordinates $x^- = z - c t$ and $\vec x_\perp$ at $\tau = z + c t
= 0$. The expansion of bound state hadronic systems in terms of
Fock states provides an exact representation of the local matrix
elements used for calculating form factors, distribution
amplitudes, and generalized parton
distributions~\cite{Brodsky:1997de}. In terms of the hadron
four-momentum $P = (P^+, P^-, \vec P_{\perp})$ with $P^{\pm} = P^0
\pm P^3$, the light-front frame independent Hamiltonian for a
hadronic composite system
$H_{LC}^{QCD} = P_\mu P^\mu  = P^-P^+ - \vec P^2_\perp$,
has eigenvalues given in terms of the eigenmass ${\cal
M}$ squared corresponding to the mass spectrum of the
color-singlet states in QCD, $H_{LC}^{QCD} \vert \Psi_h \rangle =
{\cal M}_h^2 ~ \vert \Psi_h \rangle$. The hadron state $\vert
\Psi_h \rangle$  is expanded in a Fock-state complete basis of
non-interacting $n$-particle states $\vert n \rangle$ with an
infinite number of components
\begin{equation}
\left\vert \Psi_h(P^+, \vec P_\perp) \right\rangle =
\sum_{n, \lambda_i} \int ~[d x_i~d^2 \vec k_{\perp i}]
~\psi_{n/h} (x_i, \vec k_{\perp i}, \lambda_i)
~\vert n: x_i P^+, x_i \vec P_\perp + \vec k_{\perp i}, \lambda_i \rangle ,
\label{eq:LFWFexp}
\end{equation}
where the coefficients of the light-front Fock expansion
\begin{equation}
\psi_{n/h}(x_i, \vec k_{\perp i}, \lambda_i) = \VEV{n: x_i,
\vec k_{\perp i}, \lambda_i \vert \psi_h} ,
\end{equation}
depend only on the relative partonic coordinates, the longitudinal
momentum fraction $x_i = k_i^+/P^+$, ~$\sum_{i=1}^n x_i = 1$, the
relative transverse momentum $\vec k_{\perp i}$, ~$\sum_{i=1}^n
\vec k_{\perp i} = \vec 0$, and $\lambda_i$, the projection of the
constituents' spin along the $z$ direction. The amplitudes
$\psi_{n/h}$ represent the probability amplitudes to find
on-mass-shell constituents $i$ with momentum $ x_i P^+$ and $x_i
\vec P_\perp + \vec k_{\perp i}$ and spin projection $\lambda_i$
in the hadron $h$. The measure of the constituents' phase-space
momentum integration $[d x_i~d^2 \vec k_{\perp i}]$ depends on the
normalization chosen. The complete basis of Fock-states $\vert n
\rangle$ is constructed by applying free-field creation operators
to the vacuum state $\vert 0 \rangle$ which has no particle
content, $P^+ \vert 0 \rangle =0$, $\vec P_\perp \vert 0 \rangle =
0$. Since all the quanta have positive $k^+$, the vacuum state is
unique and equal to the nonperturbative vacuum.  A one-particle
state is defined by $ \vert q \rangle = \sqrt{2 q^+} ~a^\dagger(q)
\vert 0 \rangle$ so that its normalization has the Lorentz
invariant form $\langle q | q' \rangle = 2  {q}^+ ~ (2 \pi)^3  ~
\delta ({q}^+ - {q'}^+) ~ \delta^{(2)} (\vec q_{\perp} - \vec
q_{\perp}{'})$. The measure of the phase space integration is
defined by
\begin{equation}
[d x_i~d^2 \vec k_{\perp i}] = (16 \pi^3) ~\delta\left(1 -
\sum_{j=1}^n x_j\right)~\delta^{(2)} \left(\sum_{\ell=1}^n\vec
k_{\perp \ell}\right) \prod_{i=1}^n \frac{dx_i}{x_i}~ \frac{d^2
\vec k_{\perp i}}{16 \pi^3} , \label{eq:measure}
\end{equation}
and a normalized  hadronic state
 $\VEV {\psi \vert \psi} = 1$, can be expressed as
a sum of overlap integrals of light-front wavefunctions
\begin{equation}
\sum_n  \int [d x_i~d^2 \vec k_{\perp i}]
~\vert \psi_{n/h}(x_i, \vec k_{\perp i}, \lambda_i) \vert^2 = 1 .
\label{eq:us}
\end{equation}

If the light-front
wavefunctions do not fall quickly enough, infinities appear in the
unitarity sum given by (\ref{eq:us}). To avoid this problem we
truncate the Fock-states for light-front transverse momenta $\vec
k^2_\perp$ above an ultraviolet scale
$\Lambda^2$~\cite{Lepage:1980fj}. Certainly the introduction of a
cutoff explicitly breaks the conformal invariance of the theory. When the coupling
falls as an inverse logarithm, the anomalous dimensions of an operator yield
logarithmic corrections to the scattering amplitudes; if the coupling is constant the anomalous
dimensions lead to power-law corrections~\cite{Kogut:ni}.

The cutoff is usually taken in
the limit $\Lambda \to
\infty$ when performing calculations. In practice, the cutoff has no effect on the
results provided that $\Lambda$ is much greater than all mass scales, so
calculations are carried out with a finite cutoff by defining the
wavefunctions, masses and couplings at the scale  $\Lambda$. In a
field theory where only a single scale $Q$ is relevant,  it is
natural to take the cutoff $\Lambda \sim Q$ and redefine the basic
parameters and wavefunctions at the scale $Q$. Discarding Fock
states with transverse momenta above $\Lambda$ and taking the
effective cutoff $\Lambda$ as the scale $Q$ for large values of
the momentum transfer, we can determine the large $\vec k_{\perp}^2$
momentum dependence of the light front amplitudes since we
know the conformal behavior of the bulk state as we approach the
boundary from the interior of AdS space. The correspondence
at large-$r$ follows from the conformal isometry which determines the
scaling of the string states.

To establish the correspondence between the hadron state and its dual string state and
determine the precise counting rule of each Fock component, consider an operator
$\Psi^{(n)}_h$ which creates an $n$-partonic state by applying
n-times $a^\dagger(k^+,\vec k_{\perp })$ to the vacuum state,
creating $n$-constituent individual states with plus momentum
$k^+$ and transverse momentum $\vec k_\perp$. Each higher
$n$-parton projection in the Fock expansion (\ref{eq:LFWFexp}) corresponds to the
higher-twist interpolating
operator $\Psi_h^{(n)}$. Integrating over the
relative coordinates $x_i$ and $\vec k_{\perp i}$ for each
constituent using the expression for the phase space
(\ref{eq:measure}), we find the ultraviolet behavior of
$\Psi^{(n)}_h$
\begin{equation}
\Psi^{(n)}_h(Q) \sim \int^{Q^2} [d^2 \vec k_{\perp }]^{n-1}
[a^\dagger(\vec k_\perp )]^n ~\psi_{n/h}( \vec k_{\perp }),
\label{eq:LFuv}
\end{equation}
where the  operator $a^\dagger(\vec k_{\perp })$ scales as
$1/ k_{\perp }$ at large $\vec k^2_\perp$. We obtain the scaling behavior of each
hadronic interpolating
field $\Psi^{(n)}$ from (\ref{eq:scalarAdS}) replacing the hadronic invariant mass  ${\cal M}$
with the off-shell invariant mass of the $n$-partonic state
\begin{equation}
{\cal M}_n^2 = \sum_{i=1}^n k_i^\mu k_{i \mu}
= \sum_{i=1}^n {k^2_{\perp i} + m_i^2\over x_i},
\end{equation}
since Fock-states
are eigenstates of the free light-front Hamiltonian, and ${\cal M}_n$ is
computed from the sum of the non-interacting contributions.
Near the boundary of AdS space $\Psi^{(n)} \sim Q^{-\Delta_n}$, since
${\cal M}_n$ is a hadronic scale. The
behavior of the light-front wavefunctions for large $\vec k_{\perp}^2$
then follows immediately
\begin{equation}
\psi_{n/h}(k_\perp) \sim \left( k_\perp \right)^{\Delta_n - n}
\left[ {1 \over k_\perp^2} \right]^{\Delta_n - 1}.
\label{eq:LFWFDelta}
\end{equation}

The conformal dimension
$\Delta_n$ is nominally the number of constituents
since each interpolating fermion and gauge field operator has a
minimum twist (dimension minus spin) of one.
We thus identify $\Delta_n =
n + 2\delta_n$ where $\delta_n$ represents the effect of the anomalous
dimension   of the higher-twist interpolating operators.
Thus the light-front Fock state has the leading power behavior
\begin{equation}
\psi_{n/h}(\vec k_{\perp }) \to   \left( \frac{1}{\vec k_{\perp }^2}
\right)^{n+\delta_n-1}.
\end{equation}
Notice that the light-front amplitude
$\psi_{n/h}$ does not scale as its mass dimension $M^{-n + 1}$.

Although we have derived the scaling behavior of light-front
amplitudes for a spinless hadron, identical scaling is obtained for a
spin-$\half$ hadron corresponding to a massless
dilatino state $\chi$ in the bulk that obeys the massless
Dirac equation in the 10-dimensional curved
geometry of spacetime.
In the large-$r$ conformal region the
dilatino field is expanded in terms of eigenfunctions $\eta_l$ of the Dirac
operator in compact space $X$ with eigenvalues $\lambda_l$. Specifically
$\chi(x, r, y) = \sum_l \Psi_l(x, r) \eta_l(y)$, where $\Psi$ is a spinor
field defined in a conformal d+1 space and $\eta$ is a compact
manifold spinor. For each eigenvalue $\lambda_l$ on $X$,
$\Psi_l$ obeys a  d + 1
dimensional Dirac equation in the warped $AdS_{d+1}$ space
with the solution~\cite{Henningson:1998cd}
\begin{equation}
\Psi(x,r)  = C e^{-i P \cdot x} r^{-\frac{d + 1}{2}}  \left[
  J_\alpha(\frac{{\cal M} R^2}{r})~ \mu_+(P)
+ J_{\alpha +1} (\frac{{\cal M}  R^2}{r})~ \mu_-(P) \right] ,
\label{eq:DiracAdS}
\end{equation}
where $\alpha = \lambda R - \half$, ${\cal M} = \sqrt{P_\mu P^\mu}$ and
$\hat \Gamma \mu_\pm = \pm \mu_\pm$. If $d$ is even $\hat \Gamma = i \Gamma_1 ...\Gamma_d$
is the $d$-dimensional chirality operator
and the spinors in Eq. (\ref{eq:DiracAdS}) are chiral. For $AdS_5$, $\hat \Gamma$ is the
four dimensional chirality operator $\gamma_5$.
The dominant contribution at large-$r$ also scales as $r^{- \Delta}$ where
\begin{equation}
\Delta = {d \over 2} + \vert \lambda R \vert
\label{eq:DeltaB}
\end{equation}
is the conformal dimension for a spinor field in $AdS_{d+1}$. The scaling is not
changed by the spinors since they are boundary spinors.

In QCD each Fock state is an eigenfunction of the total angular
momentum projection $J^z = \sum_{i=1}^n s_i^z + \sum_{i=1}^{n-1}
l_i^z$, where the sum over the spin $s_i^z$ corresponds to the
intrinsic spins of the n-constituent Fock states, and the sum over
the $n - 1$ orbital angular momenta
$l^z_i = -i \left( k^1_i {\partial \over \partial k^2_i}
                 - k^2_i {\partial \over \partial k^1_i} \right)$
excludes the contribution
from the motion of the center-of-mass~\cite{Brodsky:2000ii}.
Parton orbital angular momentum components in the light-front expansion of hadronic
states are essential in order to describe hadron spin-flip amplitudes in QCD, such as
the $\ell = 1$ proton $|uud\rangle$ state which is required to have a non-zero
Pauli form factor.

In $d$-dimensional Minkowski space it is simple to extend our previous scaling results
to obtain the power fall-off behavior at high relative
transfer momentum of
light-front wave functions including orbital angular momentum.
The derivation follows immediately from the
algebra of the generators of the conformal group $SO(2,d)$. In particular,
the generator of scaling transformations $D$ and the generator of
translations $P_\mu$ obey the commutation relations $[D, P_\mu] =
- i P_\mu .$ Consider a state with orbital angular momentum
$\ell$, $\Psi_\ell \sim p^\ell~\Psi$, where the interpolating field $\Psi$ is an
eigenfunction of the scaling operator $D$ with eigenvalue $ - i
\Delta$~\cite{Mack:rr}
\begin{equation}
[D, \Psi] = i ( - \Delta + x^\mu \partial_\mu) \Psi.
\end{equation}
Using the commutation relation of the operator $D$ with $P_\mu$ as
written above, we obtain the scaling dimension $\Delta_\ell$ of
the field $\Psi_\ell$ representing a  state with orbital angular
momentum $\ell$
\begin{equation}
[D, \Psi_\ell] = i ( - \Delta_\ell + x^\mu \partial_\mu) \Psi_\ell,
\end{equation}
with $\Delta_\ell = \Delta + \ell$. The dimensional scaling law
for the light-front wavefunctions at large $Q$ including
constituent orbital angular momentum $\ell$ follows directly from
the identification $\Delta_n \to \Delta_{n,\ell} = n + 2\delta_n + \ell$ in
(\ref{eq:LFWFDelta})
\begin{equation}
\psi_{n/h}(\vec k_{\perp })  \to   \left( k_{\perp } \right)^{ \ell}
\left[\frac{1}{\vec k_{\perp }^2 } \right] ^{n + \delta_n+ \ell - 1}.
\label{eq:lfwfQ}
\end{equation}

Since $O(d+1,1)$ the isometry group of $AdS_{d+1}$ in $d + 1$ dimensions,
acts as the conformal group
$SO(2,d)$ in d-dimensional Minkowski space~\cite{ref:AdS_M}, it is natural
to ask if a correspondence can be established
between a string moving in AdS space and the QCD Fock-states with orbital angular momentum.
According to the AdS/CFT duality, we would expect that all
the states of the hadronic expansion in a complete Fock-basis spanning all the Hilbert space
of the boundary theory
are matched with the string degrees of freedom in a one-to-one correspondence.
We know from string theory that the correspondence could not be established for spin
greater than
two, and a Fock-state can have high orbital angular momentum. However, the QCD eigenstate
itself has no large spin, since the $J_z$ component of each Fock-state is identical to that
of the hadron itself. If the sum of orbital components $l_i^z$ is large, it is compensated
by the sum of the constituents' spin $s_i^z$ for each of the $n-1$ orbital angular momentum
states corresponding to a Fock-state with $n$ partons.

From (\ref{eq:DeltaM}) or (\ref{eq:DeltaB}) it is clear that only if the
compact space $X$ has dimensions of order $R$, would the product $\lambda R$ be independent
of the 't Hooft coupling.
The KK states should also be protected from quantum or string effects which potentially
can give contributions of the order of the string scale to $\Delta$.
As an example, for a string that lives on $M^4 \times K$, the product of Minkowski spacetime
and a six-dimensional compact space $K$, excited KK states are non-chiral~\cite{Witten:1983ux}.
Consequently, no mechanism can in this case
prevent the excited eigenmodes from acquiring a very large mass from invariant
mass terms at tree level in the
Lagrangian. The non-chiral states acquire a mass given by the highest
scale available,
$\lambda \sim 1/ \sqrt{\alpha'}$, and decouple from the theory according to the
survival hypothesis~\cite{Georgi:1979md}, leaving effectively the massless modes.
If $ (\lambda R)^2 \sim (g_s N_C)^{1/2}$, we would expect
from (\ref{eq:DeltaM}) or (\ref{eq:DeltaB}) the  dimension of
an excited KK state to grow as $\Delta \sim (g_s N_C)^{1/4}$
at large 't Hooft coupling. In the
case of the supersymmetric Yang-Mills correspondence with Type IIB strings on
$AdS_5 \times S^5$~\cite{Maldacena:1997re}, all Kaluza-Klein excited states transform
in short supergravity multiplets and the radius of curvature of $S^5$
is also $R$. The associated dimensions are protected by the supersymmetry
algebra~\cite{Aharony:1999ti}.

Strings moving in the warped geometry of $AdS_5 \times X$  spacetime involve both
positive and negative chirality solutions as given by Eq. (\ref{eq:DiracAdS}).
In this case, the opposite chirality states have different
dimensions and thus very different dependences on $r$; they do not pair to form an
invariant mass term. Indeed only the
positive chirality solution is dominant at large $r$, corresponding to a QCD state with
proper near-conformal scaling at the AdS boundary. Contrary to strings on $M^4 \times K$,
the dimensions associated with KK modes for strings on $AdS_5 \times X$
space are protected, i.e. the dimensions
are independent of the 't Hooft coupling as long as the compact space has dimensions of order $R$.

Harmonic KK modes are eigenstates of the Laplacian on an N-sphere $S^N$
with eigenvalues $l(l + N - 1)$, where $l$ is an integer.
For $N = d + 1$, the product $\lambda R$ is thus quantized according to
$(\lambda R)^2 = l (l + d)$.
From (\ref{eq:DeltaM}) it follows that $\Delta_l = \Delta
+ l$, where $\Delta$ is the conformal dimension
corresponding to a KK zero mode. The matching of states in the
AdS/CFT correspondence is carried out near the boundary of AdS space
$r = R^2/z$ with $z \sim 0.$
We then can identify $l = \ell$.

The case for the spinor field is slightly more complex. The Dirac  operator on an N-sphere
has eigenvalues~\cite{Camporesi:1995fb} $\pm\left(l + {N\over 2}\right)$, where $l$ is a
non negative
integer $l = 0, 1, 2, ...$ Consequently $\lambda R$ obeys the quantization
condition $\lambda R = \pm \left(l + {d\over 2} + \half \right)$ for $N = d + 1$,
and the lowest KK Dirac mode on the N-sphere has non-zero mass. From (\ref{eq:DeltaB}) we obtain
for the conformal dimension of the spinor field $\Delta_l = d + l + \half$, and thus
the relation $\Delta_l = \Delta + l$ is also valid in this case.
Here also $\Delta$ is the conformal dimension for $l = 0$.
We have thus established a correspondence
of orbital angular momentum components for the Fock expansion
of meson and baryon states with the corresponding KK modes in
the compact space. Only 10-dimensional string
excitations, or stringy excitations, have masses of order of the string scale
$(\alpha')^{-1/2}$ and dimensions $\Delta \sim (g_s N_C)^{1/4}$.

To specify the angular momentum properties of hadronic
states~\cite{Karmanov:fv} in QCD, the orbital angular momentum component
of the hadron wavefunction is constructed in terms of the $n-1$
transverse momenta of the components and has the general
structure~\cite{Ji:bw}: $(k_{1 \perp}^\pm)^{\vert l_{z 1}\vert}
(k_{2 \perp}^\pm)^{\vert l_{z 2}\vert} ...(k_{(n-1)
\perp}^\pm)^{\vert l_{z(n-1)}\vert}$, with $k_{i\perp}^\pm = k_i^1
\pm i k_i^2$.
In the regime of moderate value of the 't Hooft coupling we expect
the anomalous dimension to be small
to retain the near-conformal features of QCD phenomena.
The phenomenological success of dimensional counting rules
suggest indeed that the effect of the anomalous dimensions is small. We thus write
for the hard component of the light-front wavefunction
\begin{equation}
\psi_{n/h} (x_i, \vec k_{\perp i} , \lambda_i, l_{z i}) \sim
\frac{(g_s~N_C)^{\half (n-1)}} {\sqrt{\cal N}_C}
~\prod_{i =1}^{n - 1}
(k_{i \perp}^\pm)^{\vert l_{z i}\vert} ~
\left[\frac{ \Lambda_o}{
 {\cal M}^2 - \sum _i\frac{\vec k_{\perp i}^2 + m_i^2}{x_i} + \Lambda_o^2}
 \right] ^{n +\vert l_z \vert -1}, \label{eq:lfwfN}
\end{equation}
where $\Lambda_o$ represents the basic QCD mass scale and
the normalization factor $({\cal N_C})^{- 1/2}$ depends on the color structure
of each Fock state. For example for a valence meson state ${\cal N}_C = N_C$,
for a valence baryon state ${\cal N}_C = N_C !$, and for the lowest component glue state
${\cal N}_C = N_C^2$. The form
(\ref{eq:lfwfN}) is compatible with the scaling properties
predicted by the AdS/CFT correspondence (\ref{eq:lfwfQ}) including
orbital angular momentum~\cite{ref:pwf}.

Fixed-angle large transverse momentum exclusive collision
processes in QCD  take place in the large  conformal region of AdS
space. Let us review first the results for hard meson scattering
in a theory with
gauge symmetry $SU(N_C)$ for large $N_C$ ~\cite{Coleman:1980nk}.
We use the 't Hooft double-line
representation~\cite{'tHooft:1973jz} of Feynman diagrams where a
quark propagator is represented by a single-index line and a gluon
propagator by two-index lines.
To obtain the $1/N_C$ expansion for a meson form factor $F$,
we note that there is a factor of $N_C$ from a closed color quark
loop where the photon is attached and a normalization factor
of $1 / \sqrt N_C$ for each meson wave function and thus
$F \propto N_C^0$, independent of $N_C$ as it should.

To compute the meson-meson scattering amplitude $M$ in the large $N_C$ limit, we
include a factor of $N_C$ from a closed color quark loop from quark interchange
and a factor of $1 / \sqrt N_C$ for the normalization of each meson wave
function; thus $M_{QIM} \propto 1 / N_C$. The counting rule is not changed at
fixed $g^2_{QCD}~N_C$ by any number of ladder gluon exchanges between quarks
within the same meson, as would result from the iteration of the equation of
motion of the meson wavefunction. The evaluation of planar multiple gluon
exchange diagrams is also simple. In the case of two-gluon exchange in
meson-meson scattering, the index color-counting of the gluon exchange in terms
of $q \bar q$ pairs gives an additional factor $N_C$ relative to the quark
interchange diagram from the additional quark color loop, but each vertex has a
factor $1 / \sqrt N_C$ and thus $M_{2 g} \propto 1 / N_C^2$. The $1/N_C^2$
dependence is not modified by additional exchange of gluons, since each new gluon
introduces two extra vertices and one additional closed color loop which cancels
the $N_C$ factor, and the scaling is given by the wavefunction normalization;
thus $M_{n g} \propto 1 / N_C^2$.

Baryons constitute a difficult problem in the large $N_C$ limit
since they are represented by a totally antisymmetric color state
with $N_C$ different quarks, and the number of quark lines in a
given diagram grows with $N_C.$ It is expected, however, that
baryons will follow simple scaling laws in a $1/N_C$ expansion. In
his theory of baryons at large $N_C$, Witten uses graphical
methods to describe the $n$-body force and introduces a bound
state which consists of a large number of weakly interacting
particles described in the Hartree
approximation~\cite{Witten:1979kh}. In the case under discussion,
large transverse momentum hadron-hadron scattering, the problem is
simplified, since all the $N_C$ constituents ($N_C$ large but
finite) change their collinear direction in the collision process
and acquire high transverse momentum. Thus one requires the
high-$Q$ components of the hadronic wavefunction corresponding to
large-$r$ values of AdS space given by (\ref{eq:lfwfN}), with
the  normalization factor
$1 / \sqrt {N_C !}$ for each baryon wave function. Then
combinatorial factors for the quark loops are computed, giving the non-trivial
$1/N_C$ expansion with the usual index counting in the diagrams.

Consider first the $1/N_C$ expansion for a baryon form factor $F$ for an $N_C$
component fully antisymmetric color singlet state $\vert B \rangle = {1 \over
\sqrt{ N_C !}} \epsilon_{i_1, i_2, ..., i_{N_C}} \vert q_{i_1} q_{i_2} ...
q_{i_{N_C}} \rangle$, with $\langle B \vert B \rangle = 1$. To evaluate the
combinatoric factors it is useful to consider the color quark loop with the
anti-color index from the effective remaining $N_C - 1$ quarks in the baryon
wavefunction. There is a factor $N_C$ from the closed color quark loop for the
quark in the baryon attached to the high transverse momentum photon, times the
factor $(N_C - 1)!$ to account for all the possible loops from the different
permutations in the baryon wavefunction. Taking into account the normalization
factor $\sqrt{N_C !}$ for each baryon wavefunction we obtain $F \propto N_C^0$,
independent of $N_C$.

For the baryon-baryon scattering amplitude $M$ there
is a factor $N_C$ from a closed quark loop from quark interchange
and a factor $\left[ (N_C  - 1) !\right]^2$,  which counts all the loops
from the possible permutations from the two baryon wavefunctions. There
is also a factor $(1/ \sqrt {N_C !})^4$ from the baryon
wavefunction normalization. Combining the different factors we obtain
$M_{QIM} \propto 1/N_C$. The analysis of planar multiple gluon exchange
diagrams is performed along the same lines giving an additional
$1/N_C$ factor compared to the dominant quark interchange amplitudes.
Repeating the analysis for the $1/N_C$ power counting for
meson-baryon processes, or baryon-antibaryon
scattering, we obtain identical results: the quark interchange
amplitudes scale as $1/N_C$, whereas  multiple gluon
amplitudes from planar graphs scale as $1/N_C^2$. Non-planar contributions are
naturally suppressed relative to the planar diagrams and could explain
the absence of Landshoff pinch
contributions~\cite{Landshoff:ew} in the large transverse momentum
fixed $t/s$  proton-proton scattering data, where
each pair of quarks scatter independently by gluon exchange.
These are nonplanar.

We are now in a position to write general expressions for transition
amplitudes corresponding to
the hard component of the light-front wavefunction (\ref{eq:lfwfN})
obtained from  the AdS/CFT correspondence at large values of $g_s N_C$, for
arbitrary number of components in the Fock expansion
including  orbital angular momentum.
To compare with the quantum field theory
results, the 't Hooft parameter $g_s N_C$ appearing in the
light-front wavefunction (\ref{eq:lfwfN})  has to be scaled according
$\sqrt{g_s N_C} \to g_s N_C$~\cite{ref:pwf}.
The non-perturbative expression for the form factor at large transverse momentum is
\begin{equation}
F(Q^2) \sim (g^2_{QCD}~N_C)^ {n-1}
\left[ {\Lambda_o \over Q } \right]^{2n +|l_z| - 2},  ~~g^2_{QCD} \sim g_s,
\label{eq:FF}
\end{equation}
where $n$ is the number of components in the Fock state and $\vert l_z|$ is the
angular momentum component. The extra factor $\Lambda_o^{n-1}$ comes from the soft
components of the wavefunction after
integrating the exact Drell-Yan-West expression for the form factor~\cite{Drell:1969km}
over $d^2 \vec k_\perp^{n-1}$ phase space. For
$l_z = 0$ (\ref{eq:FF}) agrees with the nominal counting
perturbative QCD results. The equivalent expression for the hadronic scattering
reaction $ A + B \to C + D$ is obtained from the general structure of the quark interchange
amplitude~\cite{Gunion:1972gy}, including the appropriate number of integration loops
 \begin{equation}
 M(Q^2)_{A + B \to C + D} \sim {(g^2_{QCD}~N_C)^{\half (n-2)} \over N_C}
\left[ {\Lambda_o \over Q} \right]^{(n + |l_z| - 4)} ,
\label{eq:HH}
\end{equation}
where $n$ is the total number of constituents entering the reaction,
$n = n_A + n_B + n_C + n_D$, and $g^2_{QCD} \sim g_s$. The additional
$\Lambda_o$ factor corresponds to
the soft momentum components of the wavefunction from phase space loop integration.

In this paper we have shown how the scaling properties of the
hadronic interpolating operator in the extended AdS/CFT space-time
theory determines the scaling of light-front hadronic
wavefunctions at high relative transverse momentum.
The angular momentum dependence of the light-front wavefunctions
also follows from the AdS/CFT correspondence mapping the Fock
components into string states in $AdS \times X$ spacetime, and
identifying the partonic states with orbital angular momentum
with the Kaluza Klein eigenmodes in the compact space $X$. The
scaling predictions agree with the perturbative QCD analysis given
in Ref.~\cite{Ji:bw}, but here the analysis is performed at strong
coupling without the use of perturbation theory.
Quark interchange
is the dominant mechanism at large momentum transfer in
the $N_C \to \infty$ limit of QCD and consistent expressions are
obtained for meson and baryon scattering at large transverse momentum. The
near-conformal scaling properties of light-front wavefunctions
lead to a number of  predictions for QCD which are normally
discussed in the context of perturbation theory, such as
constituent counting scaling laws for the leading power fall-off
of form factors and hard exclusive scattering amplitudes for QCD
processes. The ratio of Pauli to Dirac baryon form factor has the
nominal asymptotic form ${F_2(Q^2) / F_1(Q^2) }\sim 1/Q^2$, modulo
logarithmic corrections, in agreement with the perturbative
results of Ref.~\cite{Belitsky:2002kj}. Our analysis can also be
extended to study the spin structure of scattering amplitudes at
large transverse momentum and other processes which are dependent
on the scaling and orbital angular momentum structure of
light-front wavefunctions.

We thank Oleg Andreev, Joe Polchinski, Helen Quinn, J\"org Raufeisen, and Matt
Strassler for helpful comments.

\vfill

\pagebreak

\end{document}